\documentclass[hyper, nofootinbib]{PoS}
\usepackage{comment}
\usepackage{amssymb,amsmath}
\usepackage{subfigure}
\usepackage{color}
\usepackage{graphicx}% Include figure files
\usepackage{enumerate}% For non-numerical items
\usepackage{bm}% bold math
\usepackage{slashed}
\newcommand{\nn}{\nonumber} 
\newcommand{\gsim}{\raisebox{-0.7ex}{$\stackrel{\textstyle >}{\sim}$ }}
\usepackage{amssymb}
\usepackage{amsmath}
\usepackage{epsfig}

\title{Few-body physics}

\ShortTitle{Few-body physics}

\author{\speaker{Ra\'ul A. Brice\~no}\\
 Thomas Jefferson National Accelerator Facility,\\ 12000 Jefferson Avenue, Newport
  News, VA 23606, USA\\
          E-mail: \email{rbriceno@jlab.org}}
 
\abstract{
Few-body hadronic observables play an essential role in a wide number of processes relevant for both particle and nuclear physics. In order for Lattice QCD to offer insight into the interpretation of few-body states, a theoretical infrastructure must be developed to map Euclidean-time correlation functions to the desired Minkowski-time few-body observables. In this talk, I review the formal challenges associated with the studies of such systems via Lattice QCD, as first introduced by Maiani and Testa, and I also review the methodology to circumvent said limitations. The first main example of the latter is the formalism by L\"uscher to analyze elastic scattering and a second is the method by Lellouch and L\"uscher to analyze weak decays. I discus recent theoretical generalizations of these frameworks that allow for the determination of scattering amplitudes, resonances, nonlocal contribution to matrix elements, and form factors below and above inelastic thresholds. Finally, I outline outstanding problems, including those that are now beginning to be addressed. }

\FullConference{The 32nd International Symposium on Lattice Field Theory\\
                 23-28 June, 2014\\
                 Columbia University New York, NY}

\begin{document}

\section{Introduction}

%%%%%%%%%%%%%%%%%%%%%%%%%%%%%%%%%%%%%%%%%%%%%%%%%%%%%%%%
\begin{figure}[b]
\begin{center}  
\subfigure[]{\includegraphics[scale=0.25]{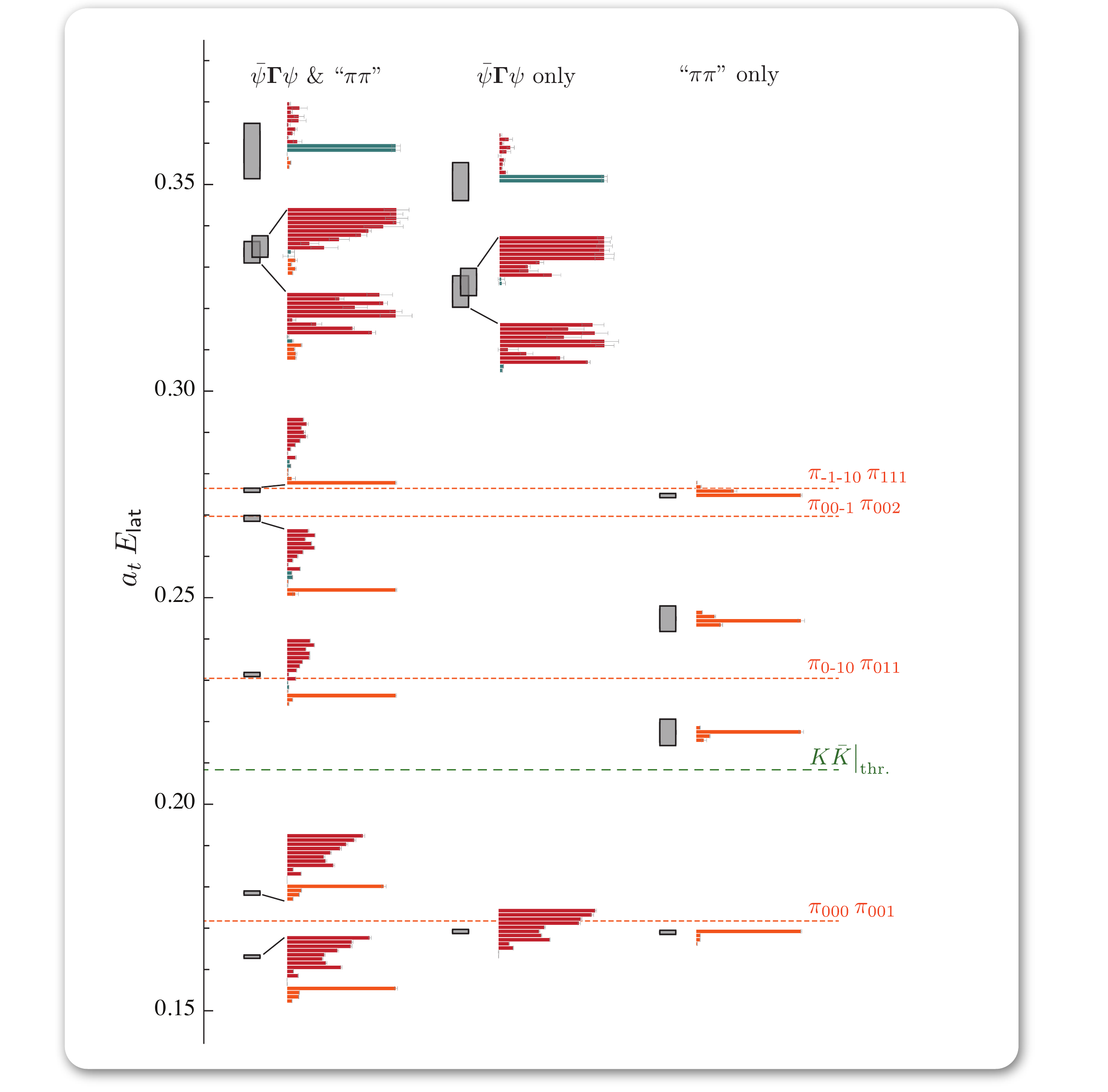}}
\subfigure[]{\includegraphics[scale=0.25]{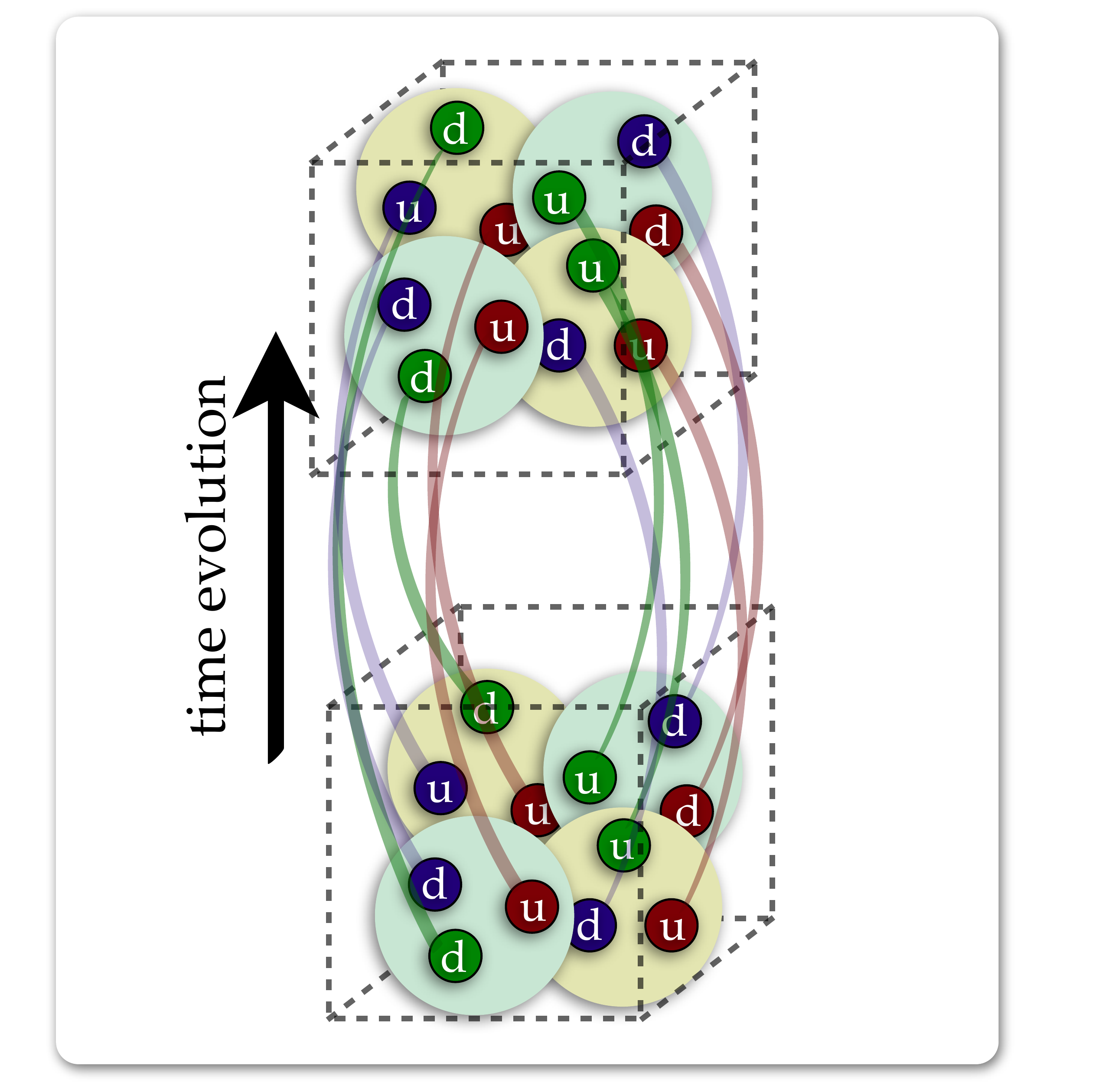}}\\
\caption{(a) Shown is the finite volume spectrum of the $A_1$ irreducible representation (irrep) of the Dic$_4$ symmetry group for $I=1$ $\pi\pi$ channel determined by the Hadron Spectrum Collaboration~\cite{Dudek:2012xn}. The rightmost column shows the spectrum determined by using only ``\emph{meson-meson-like}'' operators, the center column has the spectrum obtained only ``$q\bar{q}$\emph{-like}'' operators, and the left column shows the full spectrum obtained using a total of 22 operators including both ``\emph{meson-meson-like}'' and  ``$q\bar{q}$\emph{-like}'' operators. Reproduced with permission of Jozef Dudek. (b) For increasingly complicated systems the number of Wick contractions become prohibitively computationally expensive, e.g., the simplest $^4$He operator has naively $6!\times6!= 518,400$ contractions.  
}
\label{fig:fig1}
\end{center}
\end{figure}
%%%%%%%%%%%%%%%%%%%%%%%%%%%%%%%%%%%%%%%%%%%%%%%%%%%%%%%%
%%%%%%%%%%%%%%%%%%%%%%%%%%%%%%%%%%%%%%%%%%%%%%%%%%%%%%%%
\begin{figure}[b]
\begin{center}  
 \subfigure[]{\includegraphics[scale=.2]{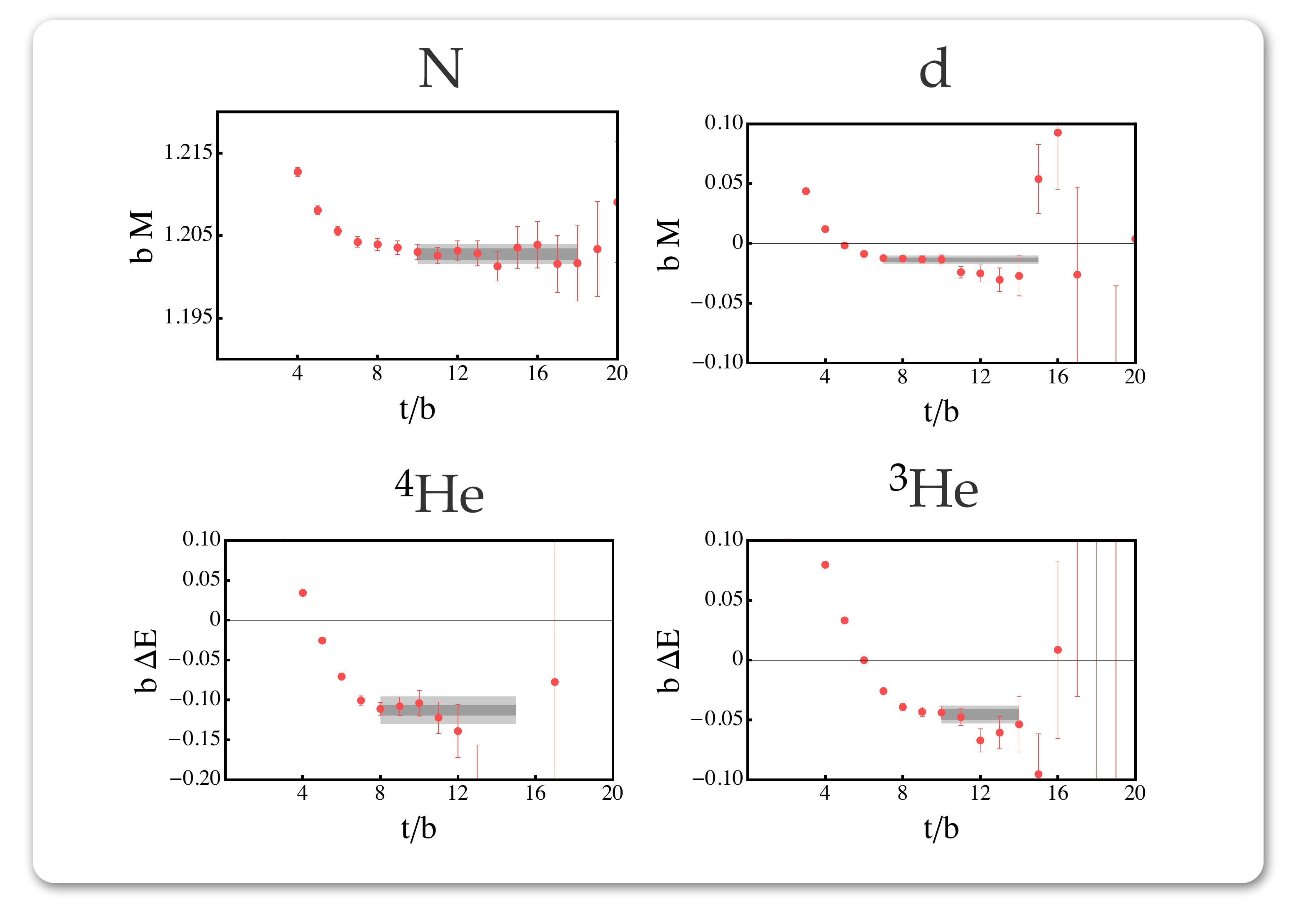}}
\subfigure[]{\includegraphics[scale=.2]{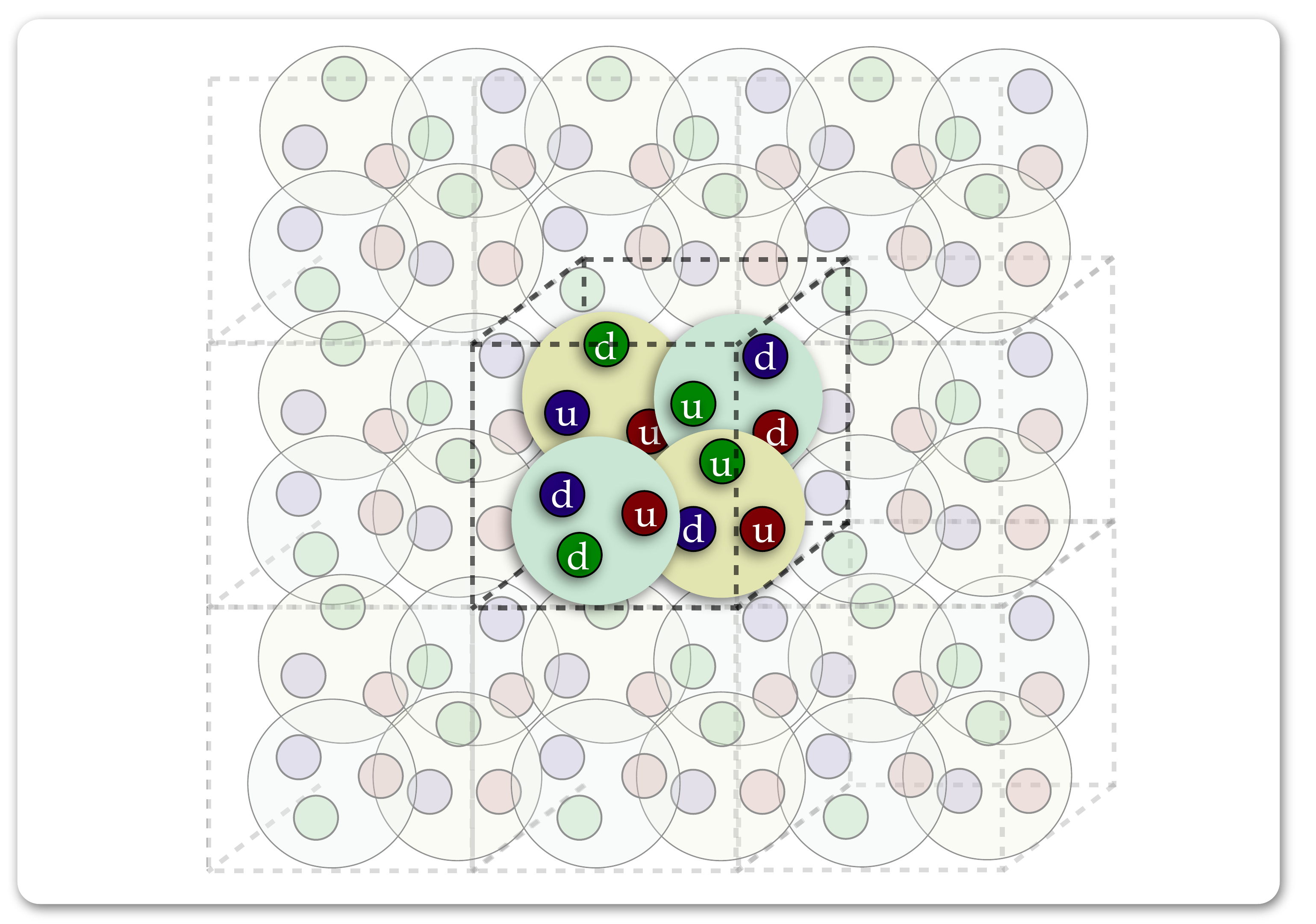}}
\\
\caption{ (a) Shown are the effective masses of the nucleon, deteuron, $^3$He and $^4$He as a function of the Euclidean time determined by the NPLQCD collaboration for $m_\pi\sim800$~MeV~\cite{Beane:2012vq}. These illustrate the deterioration of the signal as a function of the baryon number of the correlation function. Reproduced with permission of Andr\'e Walker-Lound. (b) Lattice QCD calculations are performed in a finite volume, typically with periodic (or in general twisted) boundary conditions, making the interpretation of lattice QCD observables a nontrivial one. 
}
\label{fig:fig2}
\end{center}
\end{figure}
%%%%%%%%%%%%%%%%%%%%%%%%%%%%%%%%%%%%%%%%%%%%%%%%%%%%%%%%

As the standard models continues to manifest itself as an increasingly accurate effective field theory (EFT) for the underlying building blocks of the universe, experimentalists turn their attention to rare and exotic processes. One such example is the $B^0 \rightarrow K^*\ell^+\ell^-\rightarrow K\pi \ell^+\ell^-$ semileptonic decay, which has recently be observed by the LHCb Collaboration to be in 3.7$\sigma$ discrepancy with the current Standard Model prediction~\cite{Aaij:2013qta}. This discrepancy is just as likely due to needed extensions of the Standard Model as it is due to limited understanding of the Standard Model and our ability to rigorously predict observables from it. This is a remarkably challenging task, due to primarily two reasons. First, this process depends both on the weak and strong interactions, and QCD is non-perturbative at medium to low energies. This limitation can now be circumvented by utilizing lattice QCD. This of course, requires performing numerical evaluations of QCD correlation functions in a finite Euclidean space-time. The second challenge is associated with the nature of the $K^*(892)$. The $K^*(892)$ is a resonance that can decay via the strong interaction to states with two particles ($K\pi$) or three particles ($K\pi\pi$) and it resides just below the four-particle threshold ($K\pi\pi\pi$). As a result, in order to study such a system we must understand how to relate the information gathered via lattice QCD of strongly interacting few-body, multichannel systems in a finite, Euclidean space-time to their infinite volume, Minkowski space counterparts. Although to the untrained eye this might seem to be a subtlety associated with lattice QCD calculations, it is one of the \emph{formal cornerstones} of strongly interacting few-body systems on which a sizable portion of the on-going lattice QCD efforts hinges.  

Another equally important problem that has received a great deal of attention from the lattice community is the $N^*(1440)$, ``the Roper''. From the point of view of quark models, it is unnatural that the Roper is lower in energy than the negative-parity ground state $N(1535)$. To try to shed some light onto this seeming \emph{unnaturalness}, various lattice groups have begun to determine the spin-1/2 spectrum. Although some report that the hierarchy of these two states is reversed from what is observed in nature, no calculation has dealt with the resonant nature of the Roper. This is a remarkably challenging task, since at the physical point, the Roper lies above the $N\pi$, $N\pi\pi$ and $N\pi\pi\pi$ thresholds, and it decays 30\%-40\% of the time to three particles. 

There are, arguably, four main challenges associated with studying few-body systems on the lattice. As illustrated in Figs.~\ref{fig:fig1} and ~\ref{fig:fig2}, these involve the construction of few-body operators~\cite{Detmold:2008fn, Thomas:2011rh, Basak:2005ir, Peardon:2009gh, Dudek:2010wm, Edwards:2011jj, Dudek:2012ag, Dudek:2012gj, Dudek:2012xn, Morningstar:2013bda, Lang:2013uwa} , the large number of Wick contractions~\cite{Detmold:2010au, Detmold:2013gua, Doi:2012xd, Detmold:2012eu}, the deterioration of the signal at large Euclidean time~\cite{Lepage89, Grabowska:2012ik, Endres:2011jm, Grabowska:2012ik, Detmold:2014hla, Detmold:2014rfa}, and the interpretation of the observables. In this talk I review progress made in the latter of these. This is of particular significance, since it addresses the question: \emph{``If one manages to construct the set of optimal operators, perform all the necessary Wick contractions and obtain the lattice QCD correlation functions with an unprecedented level of precision, what physical quantities are accessible from these?''}. For systems involving a single stable particle, this is a straight forward question to answer. For systems involving two particles or more this is a far more subtle matter. This is primarily due to the fact that lattice QCD calculations are performed in a finite volume. Of course, lattice QCD calculations are necessarily performed with non-zero lattice spacing and typically with unphysical quark masses. For the remainder of this discussion, we assume that the continuum limit has been taken or that discretization artifacts are well below the uncertainty of the observable of interest. Furthermore, the discussion that follows is generally independent of the quark masses. That is to say, we are interested in finding a mapping between lattice QCD observables and physical quantities for the given quark masses used in the calculation. After one obtains a physical quantities for a range of quark masses, one may proceed to extrapolate/interpolate to the physical point using either $\chi$PT or a $\chi$PT-inspired model. 

In order to find such a relation between lattice QCD observable and physical observables it is convenient to equate two of the three main representations of correlation functions. The first of these, is the spectral decomposition of the correlation function. For example, let $\mathcal{O}^\dag_{\lambda}(y_0,-\textbf{P})$ be a source operator at time $= y_0$ for a system with total momentum $\textbf{P}$ and quantum numbers defined by $\lambda$. Similarly, let $\mathcal{O}'_{\lambda'}(x_0,\textbf{P})$ be a sink operator at time $ = x_0$ for a system with total momentum $\textbf{P}$ and quantum numbers defined by $\lambda'$. The corresponding two-point function can be written (assuming the temporal extent of the lattice is infinite) by inserting a complete set of state as
\begin{eqnarray}
C(x_0-y_0)&=&\langle 0|\mathcal{O}'_{\lambda'}(x_0,\textbf{P})
\mathcal{O}^\dag_{\lambda}(y_0,-\textbf{P})|0\rangle
\\
&=&\delta_{\lambda,\lambda'}\sum_{n}e^{-E_{\lambda,n}(x_0-y_0)}
\langle 0|\mathcal{O}'_{\lambda'}(0,\textbf{P})|E_{\lambda,n};L\rangle
\langle E_{\lambda,n};L|\mathcal{O}_{\lambda}^\dag(0,-\textbf{P})|0\rangle.
\end{eqnarray}
In general, the two operators can be distinct but necessarily must have the same quantum numbers in order for the correlation function to be nonzero. $E_{\lambda,n}$ is the $nth$ finite volume energy level (the volume dependence is left implicit) for a system with quantum numbers $\lambda$, and $\langle 0|\mathcal{O}'_{\lambda'}(0,\textbf{P})|E_{\lambda,n}\rangle$ is the matrix elements of the interpolating operator and will be referred to as an overlap factor. To be able to interpret the spectrum and the overlap factors, we rely on the fact that the correlation function can be written as the sum over all Feynman diagrams of the infrared degrees of freedom (hadrons) that are allowed by the corresponding quantum numbers,
\begin{eqnarray}
C(x_0-y_0)&=&\text{\emph{Sum over all Feynman diagrams in Euclidean spacetime}}. 
\end{eqnarray}
It is this representation that lead Maiani and Testa~\cite{Maiani:1990ca} to demonstrate that the infinite volume limit of the overlap factors for systems involving two particle or more not only depend on on-shell scattering amplitudes but also on off-shell scattering amplitudes. This has been revisited in Ref.~\cite{Briceno:2014uqa, BHWL} for systems in a finite volume. Furthermore, it was this representation that lead L\"uscher to show that the finite volume spectrum can be directly related to infinite volume on-shell scattering amplitudes~\cite{Luscher:1986pf, Luscher:1990ux} and later lead Lellouch and L\"uscher to show ~\cite{Lellouch:2000pv} that $K\rightarrow\pi\pi$ decay amplitude can be obtained from the finite volume matrix element of the weak current.  \footnote{
The third representation is the path integral representation,
\begin{eqnarray}
C(x_0-y_0)&=&\frac{1}{Z_{Eucl.}}\int\mathcal{D}[U,q,\bar{q}]~\mathcal{O}'_{\lambda'}(x_0,\textbf{P})
\mathcal{O}^\dag_{\lambda}(y_0,-\textbf{P}) ~e^{-S_{\rm{Eucl.}}},
\end{eqnarray} which allows one to approximate the integral using Monte Carlo sampling and thereby numerically evaluating the correlation functions. }

%%%%%%%%%%%%%%%%%%%%%%%%%%%%%%%%%%%%%%%%%%%%%%%%%%%%%%%%
%%%%%%%%%%%%%%%%%%%%%%%%%%%%%%%%%%%%%%%%%%%%%%%%%%%%%%%%
%%%%%%%%%%%%%%%%%%%%%%%%%%%%%%%%%%%%%%%%%%%%%%%%%%%%%%%%
\begin{comment}Having reached an unprecedented level of success in the determination of one-particle observables, the lattice QCD community has turned its attention to increasingly challenging yet interesting physical systems. 

In this talk I review the challenges associated with studying systems involving two particles or more, and I give a progress report on the current status of the field. 

Over half a century ago, Quantum Chromodynamics (QCD) was recognized as the fundamental theory of the strong nuclear force, and its connection to experimentally determined observables continues to be a source of research at the lab and across the globe. 
\end{comment}

\section{The finite volume spectrum and scattering amplitudes}
Although the observation that scattering amplitudes cannot be directly determined from the amplitude of the two-point correlation functions determined from lattice QCD is commonly attributed to Maiani and Testa's \emph{no-go theorem}, this is a problem whose solutions predates it. By the time that Maiani and Testa made this observation, Martin L\"uscher had already found a non-peturbative mapping between the finite volume spectrum and the infinite volume scattering amplitude. \footnote{L\"uscher was not the first to observe the fact that the finite volume spectrum and the infinite volume S-matrix are relatable, but he was the first to find a non-perturbative solution for said problem.} In his original work, L\"uscher found a master equation for the spectrum of two particles with zero spin at rest. This has been generalized for various two-body systems with nonzero total momenta, e.g., see Refs.~\cite{Luscher:1986pf, Luscher:1990ux, Kim:2005gf, Christ:2005gi, Rummukainen:1995vs, Bernard:2008ax, Hansen:2012tf, Briceno:2012yi}. Earlier this year it was shown that generalizing this work to two-particle systems with arbitrary intrinsic spin, any number of open two-particle channels, nonzero total momentum, and generic twisted boundary conditions on a volume shaped as a generic rectangular prism  is straightforward~\cite{Briceno:2014oea}. Generalizing this idea to systems with energies above the three-particle threshold is remarkably challenging~\cite{Polejaeva:2012ut, Briceno:2012rv, Hansen:2014lya, Hansen:2014eka, Hansen:2014lya, Guo:2013qla, Kreuzer:2010ti}. Here we discuss process made this past year on these two fronts.

%%%%%%%%%%%%%%%%%%%%%%%%%%%%%%%%%%%%%%%%%%%%%%%%%%
%\begin{figure*}[t]\begin{center}\subfigure[]{\label{fig:FVcorr}\includegraphics[scale=0.40]{./figures/FVcorr2}}\\\subfigure[]{\label{fig:kernel}\includegraphics[scale=0.20]{./figures/kernel}}\subfigure[]{\label{fig:1bodyprop}\includegraphics[scale=0.20]{./figures/1bodyprop}}\caption{a) Shown is the definition of the finite volume two-particle correlation function. The solid lines denote two-particles in the $``1"$ channel, dashed lines denote particle in the $``2"$ channel. The correlation function is written in terms of the c.m.~kernel, ${K}^*$, and the fully dressed single particle propagators. b) Shown is ${K}^*$ for the first channel, which is the sum of all two-particle irreducible s-channel diagrams. Explicitly shown are examples of diagrams that are included in the kernel: contact interactions, t- and u-channel diagrams. In general, all diagrams allowed by the underlying theory where the intermediate particles cannot all simultaneously go on-shell are absorbed into the kernel. As described in the text, in this study we are restricted to energies where only two-particle states are allowed to go on-shell. c) Shown is the definition of the fully dressed one particle propagator in terms of the one particle irreducible (1PI) diagrams.   }\label{fig:corr2}\end{center}\end{figure*}

%%%%%%%%%%%%%%%%%%%%%%%%%%%%%%%%%%%%%%%%%%%%%%%%%%
\begin{figure*}[t]\begin{center}
{\label{fig:FVcorr}\includegraphics[scale=0.35]{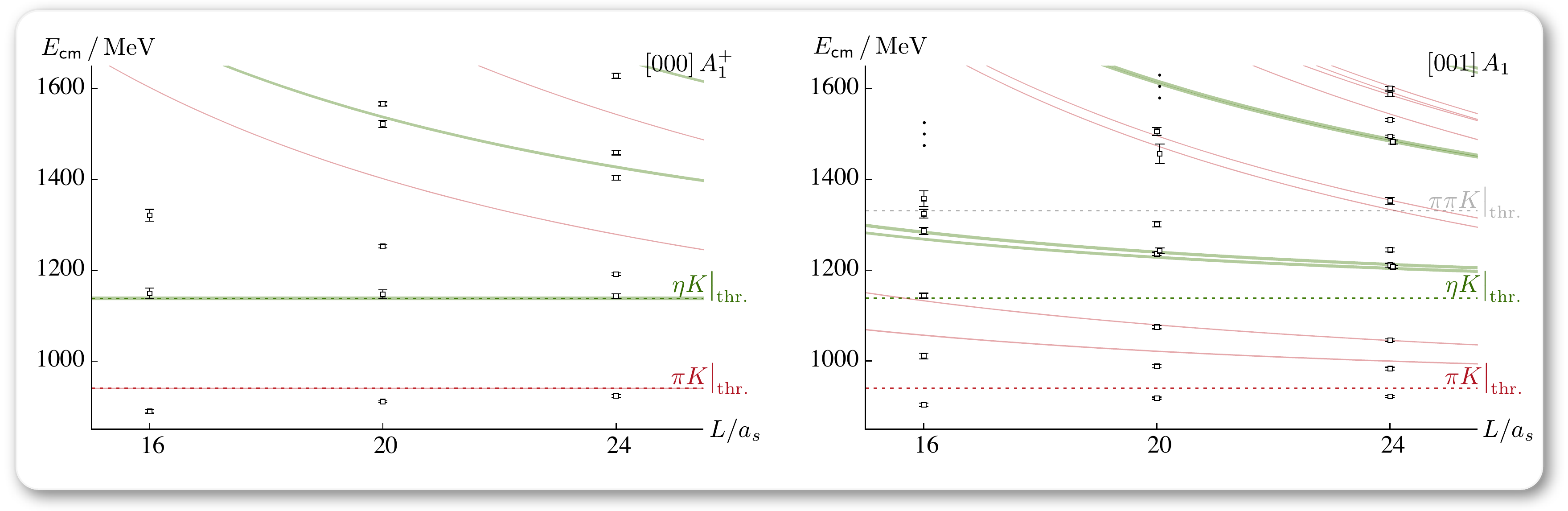}}\\
\caption{
Shown are two examples of the $K\pi-K\eta$ coupled channel spectrum as a function of the spatial extent $L=16, 20,$ and $24$ determined by the Hadron Spectrum Collaboration [HadSpec] using $m_\pi\sim$390 MeV~\cite{Dudek:2014qha}. On the left, the determined center of mass spectrum (error bars) for the $A_1^+$ irrep of the Octahedral group is compared with the spectrum in the absence of interactions (red line for $K\pi$ and green line for $K \eta$). On the right is the same for the $A_1$ irrep of the Dic$_4$ group, corresponding to a system with total momentum $|\textbf{P}|=2\pi/L$.The figures are reproduced with the permission of David Wilson and Jozef Dudek.}\label{fig:hadspec1}\end{center}\end{figure*} 
%%%%%%%%%%%%%%%%%%%%%%%%%%%%%%%%%%%%%%%%%%%%%%%%%%
\subsection{Two-body finite volume spectrum~\label{sec:FVspec}}
In order to find the condition that the spectrum must satisfy, one need only look at the poles of the correlation function. For energies below the three-particle threshold, it is sufficient to evaluate the sum of all the $\textbf{2}\rightarrow\textbf{2}$ fully dressed Feynman diagrams in a finite volume. This amounts to replacing integrals over intermediate momenta to sum over the allowed momenta. One can obtain a model independent relation between the spectrum and the infinite volume scattering amplitude~\cite{Briceno:2014oea} %
\begin{eqnarray}
\!\!\!\!\!\!\!\!\!\!
\det~[\mathcal{M}_2^{-1}+\delta \mathcal{G}^V]~\equiv~{\det}_{\rm{oc}}\left[{\det}_{{pw}}~[\mathcal{M}_2^{-1}+\delta \mathcal{G}^V]\right]
=~{\det}_{\rm{oc}}\left[{\det}_{{pw}}~[\mathcal{K}_2^{-1}+F_2]\right]=0,
\label{eq:QC}
\end{eqnarray}
where the determinant ${\det}_{\rm{oc}}$ is over N open coupled channels and the determinant $\det_{{pw}}$ is over the $|\ell S,JM_J\rangle$ basis. $\ell S$ denote the orbital angular momentum and the total spin of the two-particle systems, respectively, and $J,M_J$ is the total angular momentum and its azimuthal component. $\mathcal{M}_2$ is the on-shell c.m. scattering amplitude which in general can couple states with different $(\ell S)$ and/or different flavor-space channels, .e.g., $K\pi -K\eta$. The K-matrix, ${\cal K}_2$, is related to the scattering amplitude via $\mathcal{M}_{2}^{-1}={\cal K}_2 -i\rho$, $\rho$ is the phase space factor. The matrix $\delta \mathcal{G}$ depends on the volume, the total momentum, masses, energy and spin of the particles of interest and its matrix elements are explicitly written in Ref.~\cite{Briceno:2014oea}. Also used is the finite volume function, $F_2$, which is related to $\delta \mathcal{G}$ via $\delta \mathcal{G}=F_2+i\rho$. Prior to decomposing onto angular momentum, both ${\cal K}_2$ and $F_2$ are real. These matrix elements are written explicitly for systems in a volume that is a generic rectangular prism and that has twisted boundary conditions. Note that the majority of lattice QCD calculations use period boundary conditions on the spatial extent of the lattice, and these are a subset of twisted boundary conditions. 

This result is model-independent, non-perturbative and universal. This is due to the fact that in arriving at Eq.~\ref{eq:QC}, one only needs to understand the analytic structure of the infinite and finite volume Feynman diagrams. No assumptions need to be made about the nature of the particles in the systems or the underlying effective field theory, other than it must satisfy unitary.  

In general, the matrix inside the determinant is infinitely large, but in practice one must truncate it. This truncation is justified by the fact that although different partial waves mix due to the reduction in symmetry, for low energies the higher partial waves are kinematically suppressed. Similarly, the contribution from states that cannot go on-shell are exponentially suppressed $\leq\mathcal{O}(e^{-m_\pi L})$. Furthermore, the matrix inside the determinant can be blocked diagonalized since different irreducible representations (irreps) of the underlying symmetry group do not couple to each other. After having performed this, one finds the condition the spectrum satisfies. 
%%%%%%%%%%%%%%%%%%%%%%%%%%%%%%%%%%%%%%%%%%%%%%%%%%
\subsubsection{Correlation functions and \emph{``off-shell''} states~\label{sec:nogotheorem}}
Having found the relationship between the finite volume spectrum and the on-shell infinite volume scattering amplitude, it is easy to explain why the overlap factors for the correlation function will, in general, depend on \emph{``off-shell''} scattering amplitudes. In order to explain this, we can consider energies where only one channel can go on-shell. Imagine the system to be composed of two pions with zero total momentum. Therefore, the simplest interpolation operator one could imagine would have the form $\mathcal{O}(t)=\pi(t,\textbf{k})\pi(t,-\textbf{k})$, where $\textbf{k}=2\pi\textbf{d}/L$ and $\textbf{d}$ is an integer triplet. Let $E_n$ be the $nth$ energy satisfying Eq.~\ref{eq:QC}. Unless the interactions are exactly zero, $E_n\neq2\sqrt{\textbf{k}^2+m_\pi^2}=2\omega_k$, therefore the free states cannot go on-shell. Instead there is be a momentum $\textbf{q}_n$ that satisfies $E_n=2\sqrt{\textbf{q}_n^2+m_\pi^2}$, corresponding to an on-shell state. Consequently, the overlap factor $|\langle 0|\mathcal{O}(0)|E_{n};L\rangle|$ must encode information regarding on-shell as well as off-shell states. Reference~\cite{Briceno:2014uqa} shows that this factor is proportional to $|\textbf{K}(E_n,\textbf{k},\textbf{q}_n)|$, which is the absolute value of the K-matrix for a incoming state with energy and momentum $(E_n,\textbf{k})$ and therefore off-shell, while the on-shell outgoing state with energy and momentum $(E_n,\textbf{q}_n)$. Furthermore, it is straightforward to shown that the overlap factor is inversely proportional to $|E_n-2\omega_k|$~\cite{Bernard:2012bi}, explaining why certain two-body operators will more strongly overlap with a given eigenstate. Following this logic, if the operator used is instead a ``$q\bar{q}$\emph{-like}'' operator, one should not expect to have a dependence on off-shell states. 
%%%%%%%%%%%%%%%%%%%%%%%%%%%%%%%%%%%%%%%%%%%%%%%%%%
\subsubsection{An example: $\pi K-\eta K$ at $m_\pi\sim$~400~MeV\label{sec:Kpietapi}}
To this day, the only implementation of Eq.~\ref{eq:QC} to understand the lattice QCD spectrum for energies where more than one two-particle channel can go on-shell is by the Hadron Spectrum Collaboration [HadSpec]~\cite{Dudek:2014qha} and was presented at the 32nd International Symposium on Lattice Field Theory by David Wilson~\footnote{The first  exploratory calculation of a  coupled-channel system in a finite volume was by Guo using a 1+1D toy model~\cite{Guo:2013vsa}.}. They determined the S, P, and D-wave scattering phase shifts and inelasticities for the  $\pi K-\eta K$ coupled-channels using light quark masses corresponding to $m_\pi\sim$~400~MeV. There were over 100 energy levels determined using three different volumes (satisfying $m_\pi L\gsim 4$) and five different types of boosts, corresponding to $\textbf{d}=\textbf{P}L/2\pi=\left\{[000],[001],[011],[111],[002]\right\}$ and all allowed cubic rotations. Figure~\ref{fig:hadspec1} shows a  sample of the energy levels determined and are compared with the non-interacting energy levels. It is important to emphasize that for the systems considered, in general the S, P and D partial waves mix onto each other, and in fitting the spectrum, the HadSpec Collaboration have taken into account these effects.

\begin{figure*}[t]\begin{center}
{\includegraphics[scale=0.35]{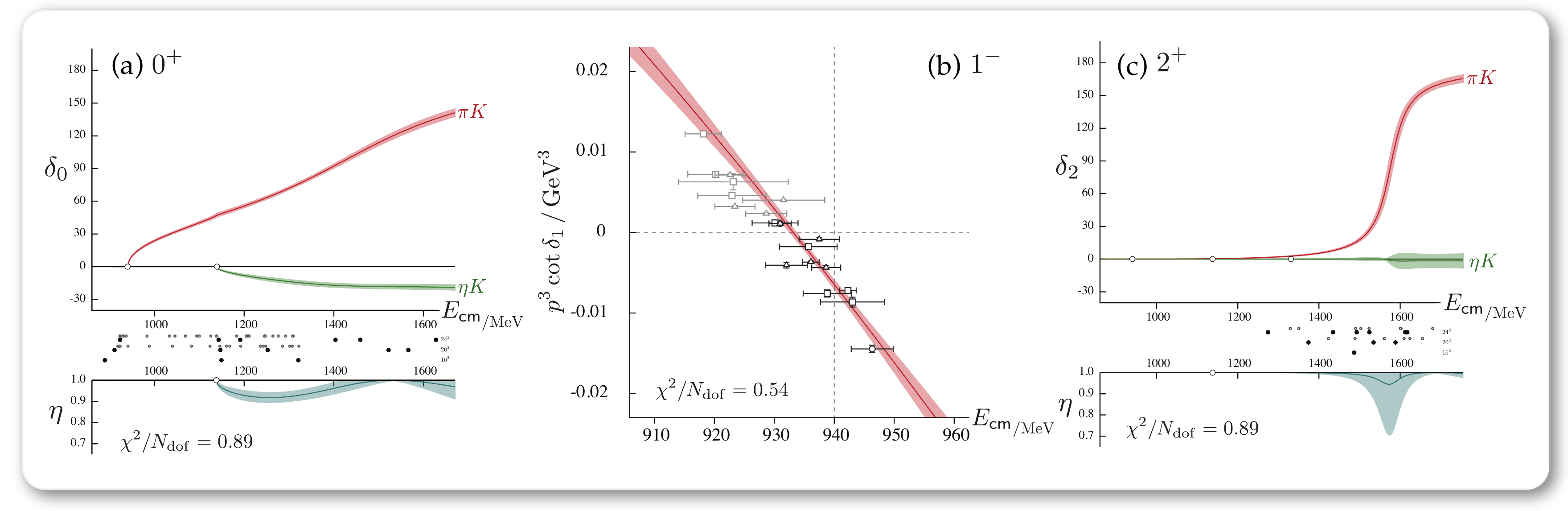}} 
\caption{
Shown are two examples of the $K\pi$ phase shifts (red), $K\eta$ phase shift (green) and the inelasticity (blue) for the (a) S, (b) P and (c) D partial waves determined by the Hadron Spectrum Collaboration as described in the text~\cite{Dudek:2014qha}. The figure is reproduce with the permission of David Wilson and Jozef Dudek.}\label{fig:hadspec2}\end{center}\end{figure*}

After determining the spectrum, the HadSpec collaboration proceeded to them fit the spectrum using Eq.~\ref{eq:QC} using a range of possible parameterizations of the scattering amplitude. Following this procedure they obtain the scattering phase shifts and inelasticities depicted in Fig.~\ref{fig:hadspec2}. By analytically continuing the scattering amplitudes onto the complex plane, in the S-wave channel, they observe a wide resonance that they associated with the kaon resonance that is known as the $K^\star_0(1430)$ at the physical point. They also find a ``\emph{virtual bound state}'', i.e., a pole on the real axis located on unphysical sheet, which they identify with the $\kappa$. In the P-wave channel, they find that the $K^\star(892)$ is bound. Most surprising of all, in the D-wave channel they find a narrow resonance, which presumably corresponds to the $K^\star_2(1430)$. This is striking since, as can be seen from Fig.~\ref{fig:hadspec2}, the energies used in fitting the resonance lies well above the $K\pi\pi$. In performing the calculation they have not accounted for three body effects in neither the operator construction nor the interpretation of the spectrum. The rather good fit of the spectrum would then lead one to postulate that at the quark masses used, following the arguments sketched in Sec.~\ref{sec:nogotheorem}, the $K\pi\pi$ only weakly couples to the two-body system.

%%%%%%%%%%%%%%%%%%%%%%%%%%%%%%%%%%%%%%%%%%%%%%%%%%
%%%%%%%%%%%%%%%%%%%%%%%%%%%%%%%%%%%%%%%%%%%%%%%%%%

\subsubsection{Including perturbative QED effects}

Currently all lattice QCD calculations involving states that have overlap with two particles or more are performed without dynamical QED effects. Nevertheless, some calculations of single particle states are beginning to incorporate QED interactions, e.g., see Refs.~\cite{Blum:2007cy, Blum:2010ym, Aoki:2012st, deDivitiis:2013xla, Borsanyi:2014jba}. The infinite range of QED in general leads to a large (power-law) FV effects. These effects can become manageable if one modifies the implementation of QED on the lattice, for instance by removing the zero mode of the photon field. In this framework, finite volume effects have been explored in several one-particle observables~\cite{Duncan:1996xy, Hayakawa:2008an, Davoudi:2014qua, Borsanyi:2014jba}. 

Most recently, Beane and Savage have shown that scattering parameters of low-lying charged two-particle states can in principle be determined by performing calculations of the finite volume spectrum with dynamical QED+QCD~\cite{Beane:2014qha}. In their work, they calculate the leading order QED effects, i.e., $\mathcal{O}(\alpha)$, to the quantization condition of two particle systems. As a result, their result does not hold for systems with QED bound states. In order to make some progress, they restrict themselves to non-relativistic systems below the QCD t-channel cut and the QED inelastic threshold (photon production). They restrict the two-particle system to be in an S-wave and ignore higher partial waves. Within these approximations, they find at ${\cal O}(\alpha)$, the two-particle finite volume spectrum satisfies
\color{black}
\begin{eqnarray}
 - {1\over a_{C,L}^\prime} \ +\ \frac{1}{2}r_{0,L}^\prime~p^2 \ +\ ...
 & = & \frac{1}{\pi L}{S}^{ C}\left[p,a_C,r_0,L\right] \ +\ {\alpha m}\left[\ln\left({4\pi\over\alpha m {\rm  L}}\right) - \gamma_E \right]  
 \ +\ ...
\ , 
\end{eqnarray}
where $p$ is the relative momentum of the two on-shell particles. $a_{C,L}^\prime$ and $r_{0,L}^\prime$ are finite volume functions that in the infinite volume limit asymptote to the scattering length and effective range, and ${S}^{ C}$ is a finite volume function that is closely related to the Zeta functions but also depends on the physical quantities listed in its argument~\cite{Beane:2014qha}. 
 
%%%%%%%%%%%%%%%%%%%%%%%%%%%%%%%%
\subsection{Three-body finite volume spectrum \label{sec:3body}}

Recently there have been attempts to generalize Eq.~\ref{eq:QC} to energies where three or more particles can go on-shell~\cite{Polejaeva:2012ut, Briceno:2012rv, Guo:2013qla, Kreuzer:2010ti}, the most recent of which was presented in this conference and is reviewed here~\cite{Hansen:2014eka, Hansen:2014lya}. In their work, Hansen and Sharpe, found a master equation describing the finite volume spectrum for three degenerate scalar bosons. In arriving at their result, they have assumed for simplicity that the two-body K-matrix, ${\cal K}_2$, is finite in the energy range considered. This amounts to restricting the phase shift of the $\ell th$ partial wave to satisfy $|\delta_{\ell}|<\pi/2$. This currently prevents the implementation for studies of, say, $\rho\pi$ scattering. Furthermore, they have assumed an underlying $Z_2$ symmetry that, for examples, prohibits $\textbf{2}\rightarrow \textbf{3}$ processes.

In order to introduce the result, it is necessary to first define the degrees of freedom of a three-particle systems. Let $(E,\textbf{P})$ be the total energy and momentum and let all the particle be on-shell. The system can be separated onto an ``\emph{spectator}'' and a ``\emph{pair}''. 
 Let $\textbf{k}$ be momentum of the spectator and $(\ell,m)$ be the angular momentum projection of the pair in its c.m. frame. The matrices and determinant below are evaluated in the space of the direct product of spectator momentum and angular momentum of the pair. Hansen and Sharpe arrive at a non-perturbative, fully relativistic result,  
\begin{equation}
\det~[F_3^{-1} + i {\cal K}_{\text{df},3} ] = 0\,.
\label{eq:QC3}
\end{equation}
${\cal K}_{\text{df},3}$ is the \emph{``divergence free''} three-body K-matrix and its relationship to the fully $\textbf{3}\rightarrow\textbf{3}$ scattering amplitude is defined in Ref.~\cite{Hansen:2014eka}. Unlike Eq.~\ref{eq:QC}, where the infinite volume and finite volume kinematic functions are nicely separated, the function $F_3$ is not a purely kinematical function; it also depends on 
${\cal K}_2$:
\begin{equation}
F_3 = \frac1{2\omega L^3}
\left[-\frac{2F}3 
+ \frac1{F^{-1} + (1+K G_{NR})^{-1} K F}\right]\,,
\label{eq:F3}
\end{equation}
where $K$ and $F$ are respectively ${\cal K}_2$ and $F_2$
multiplied by $\delta_{\vec k,\vec k'}$. The matrix $\omega L^3$ is diagonal with elements $L^3\,\sqrt{\vec k^2+m^2}$,
and $G_{NR}$ is a regulated free non-relativistic propagator defined in Ref.~\cite{Hansen:2014eka}.

In order to implement this formalism for the three-body system, one first needs to obtain an analytic parameterization of ${\cal K}_{2}$ by interpolating the values obtained using the two-particle quantization, as was done, for example, in Ref.~\cite{Dudek:2014qha} and reviewed in Sec.~\ref{sec:Kpietapi}. Consequently, $F_3$ in Eq.~\ref{eq:F3} would be also determined. After determining the three-body spectrum via lattice QCD, by inputing the spectrum and $F_3$ onto Eq.~\ref{eq:QC3} one could then constrain ${\cal K}_{\text{df},3}$. In their work, Hansen and Sharpe, in addition to giving a detailed derivation of Eq.~\ref{eq:QC3}, they also demonstrate how this determinant condition can be suitably truncated.

%%%%%%%%%%%%%%%%%%%%%%%%%%%%%%%%%%%%%%%%%%%%%%%%%%%%%%%%%%%%%%%%%%%%%
\section{Three-point correlation functions}
As discussed in the introduction, the determination of elastic and inelastic form factors via lattice QCD is essential for our understanding in a wide range of experimentally observed (and unobserved) processes. One way to obtain matrix elements of external current is to study the shift in the spectrum of a system in the presence of an external field. A recent example of this method was presented in this conference~\cite{Lee:2014iha} and it was recently implemented in the determination of magnetic moments of light nuclei via lattice QCD~\cite{Beane:2014ora}. The advantage of this methodology is that one avoids having to evaluate three-point functions, which in general are more computationally challenging and costly. That being said, the vast majority of experimentally observed hadrons are unstable under the strong interaction, and generalization of this framework for systems involving two or more particles is not straightforward, e.g., see Ref.~\cite{Detmold:2004qn}. 

Alternative to introducing an external field, one can proceed to simply evaluate a three-point correlation function involving an external current, $\mathcal{J}_{\lambda_c}(t_c,-\textbf{Q})$, inserted in time $ = t_c$ with momentum $\textbf{Q}$,
\begin{align}
C^{(3)}(x_f-t_c,t_c-x_i)&=\langle 0|\mathcal{O}'_{\lambda_f}(x_f,\textbf{P}_f)
\mathcal{J}_{\lambda_c}(t_c,-\textbf{Q})
\mathcal{O}^\dag_{\lambda_i}(y_0,-\textbf{P}_i)|0\rangle
\\
&=\sum_{n_i,n_f}e^{-E_{\lambda_f,n_f}(x_f-t_c)}e^{-E_{\lambda_i,n_i}(t_c-x_i)}
\langle 0|\mathcal{O}'_{\lambda_f}(0,\textbf{P}_f)|E_{\lambda_f,n_f};L\rangle\nn\\
&
\hspace{1.5cm}\times
\langle E_{\lambda_f,n_f};L|
\mathcal{J}_{\lambda_c}(t_c,-\textbf{Q})|E_{\lambda_i,n_i};L\rangle
\langle E_{\lambda_i,n_i};L|\mathcal{O}_{\lambda_i}^\dag(0,-\textbf{P}_i)|0\rangle.
\end{align}
Here, $\lambda_c$ labels all of the quantum numbers of the external current of interest. For systems where $E_{\lambda_i,n_i}$ and $E_{\lambda_f,n_f}$ correspond to QCD stable states, one expects the finite volume matrix elements of the external current, $\langle E_{\lambda_f,n_f};L|\mathcal{J}_{\lambda_c}(t_c,-\textbf{Q})|E_{\lambda_i,n_i};L\rangle$, to be exponentially close to corresponding infinite volume matrix element, after removing the momentum conserving delta-function. For systems where either the initial or(and) final state is(are) unstable or loosely bound these matrix elements states are expected to suffer of large finite volume effects. Therefore, prior to performing these calculations one must find a mapping between finite volume matrix elements of external current and the infinite volume matrix elements. 
%%%%%%%%%%%%%%%%%%%%%%%%%%%%%%%%%%%%%%%%%%%%%%%%%%%%%%%%%%%%%%%%%%%%%
%%%%%%%%%%%%%%%%%%%%%%%%%%%%%%%%%%%%%%%%%%%%%%%%%%%%%%%%%%%%%%%%%%%%%

\subsection{$\textbf{1}\rightarrow\textbf{2}$ transition amplitudes}
The relationship between finite volume matrix elements and infinite volume physical observables is in a formally less mature state than the interpretation of the finite volume spectrum discussed in Sec.~\ref{sec:FVspec}. The subtle relation between finite volume matrix elements and infinite volume amplitudes was first addressed by Lellouch and L\"uscher in the context of $K\rightarrow\pi\pi$ decays~\cite{Lellouch:2000pv}. Since their original work, their idea has been extended to increasingly complicated systems. Just this year, two advances were made and presented in this conference. First, Agadjanov et al. demonstrated how this framework can be implemented to study electromagnetic decays of baryonic resonances via lattice QCD with arbitrary $Q^2$, e.g., $N\gamma^*\rightarrow\Delta\rightarrow N\pi$~\cite{Agadjanov:2014kha}.

The second development was by Hansen and Walker-Loud in collaboration with the author of this review~\cite{Briceno:2014uqa, BHWL}. References~\cite{Briceno:2014uqa, BHWL} present a generic relation between matrix elements of currents that couple one-particle and two-particle states to the corresponding infinite volume transition amplitude, $\mathcal{A}$. The only assumptions made in this work is that initial(final) state only couples to one(two)-particle states and the individual particles involved have no intrinsic spin. Otherwise, the particles can carry arbitrary momenta, the final two-particle states can be in a cubic irrep, angular momenta are allowed mixed in accordance to symmetries of the system, in general any number of two-particle states can go on-shell, the volume could be any rectangular prism with arbitrary twisted boundary conditions. 

The absolute value of the matrix elements of the external current can be written in terms of the residue of the finite volume two-particle propagator at the $n_fth$ pole, $\mathcal{R}_{\lambda_f,n_f}$, and the transition amplitude evaluated at that kinematic point, $\mathcal{A}_{\lambda_f,n_f;\lambda_c}$. In general, $\mathcal{R}_{\lambda_f,n_f}$ is a matrix in the space of open channels and the in the space of angular momenta that couple to a two-particle state with quantum numbers compactly labeled here by $\lambda_f$.  $\mathcal{R}_{\lambda_f,n_f}$ depends on the spectrum and the K-matrix of the two-body systems. Having obtained the spectrum, one can determine the K-matrix using the two-body quantization conditions, Eq.~\ref{eq:QC}, thereby determining $\mathcal{R}_{\lambda_f,n_f}$.  For an explicit expression of $\mathcal{R}_{\lambda_f,n_f}$ see the aforementioned reference. Similarly, $\mathcal{A}_{\lambda_f,n_f;\lambda_c}$ is a column vector in this space. Using this notation, one can show that the matrix elements satisfy
\begin{equation}
\left|\langle E_{\lambda_f,n_f};L|\mathcal{J}_{\lambda_c}(0,\textbf{P}_f-\textbf{P}_i)| E_{\lambda_i,0};L\rangle\right| =
\sqrt{ \frac{\mathcal{N}_i~\mathcal{N}_f}{2E_{\lambda_i,0}}}
\sqrt{
\left[\mathcal{A}^\dagger_{\lambda_f,n_f;\lambda_c}~\mathcal R_{\lambda_f,n_f}~\mathcal{A}_{\lambda_f,n_f;\lambda_c}\right]
},
\label{eq:matJaieps}
\end{equation}
where $\mathcal{N}_i$ and $\mathcal{N}_f$ are the normalization of the initial and final states, respectively, in a finite volume. References~\cite{Briceno:2014uqa} chose to set these to one, but another common choice in the literature is to set these equal to twice the respective energy. This expression allows one to determine the absolute value of the transition amplitude. The strong phase associated with this amplitude can be directly determined from the spectrum, using Eq.~\ref{eq:QC}. Furthermore, the overall sign of the given transition amplitude is not a physical observable, but the relative sign between different current is a physical observable. One can also show that the ratio of matrix element of two distinct current, ${\mathcal{J}}_{\lambda_c,1}$ and ${\mathcal{J}}_{\lambda_c,2}$, satisfies 
\begin{align} 
\frac{\langle E_{\lambda_f,n_f}\textbf{P}_f;L|{\mathcal{J}}_{\lambda_c,1}(0,\textbf{P}_f-\textbf{P}_i)| E_{\lambda_i,0}\textbf{P}_i;L\rangle}{\langle E_{\lambda_f,n_f}\textbf{P}_f;L|{\mathcal{J}}_{\lambda_c',2}(0,\textbf{P}_f-\textbf{P}_i)| E_{\lambda_i,0}\textbf{P}_i;L\rangle} =
\frac{\mathcal{A}_{\lambda_f,n_f;\lambda_c',2}^\dag~\mathcal{R}_{\lambda_f,n_f}~\mathcal{A}_{\lambda_f,n_f;\lambda_c,1}}
{\mathcal{A}_{\lambda_f,n_f;\lambda_c',2}^\dag~\mathcal{R}_{\lambda_f,n_f}~\mathcal{A}_{\lambda_f,n_f;\lambda_c',2}}.
\label{eq:ratio}
\end{align} 
From this it is clear that the relative sign between the matrix elements can help constrain the relative sign between the transition amplitudes.

To this day, the only implementation of this formalism has been towards the study of $K\rightarrow\pi\pi$ decays~(e.g., see Refs.~\cite{Ishizuka:2014nfa, Blum:2012uk, Boyle:2012ys, Blum:2011pu, Blum:2011ng}). Lellouch and L\"uscher first proved this was possible assuming the final state is at rest. This restriction was later removed~\cite{Kim:2005gf, Christ:2005gi}. This was then extended to allow for two open channels in the final state, as in the case, for example, $D\rightarrow\pi\pi$~\cite{Hansen:2012tf}. The result above compactly summarizes all of these scenarios, and it opens up the possibility of a large number of new calculations discussed in Ref.~\cite{Briceno:2014uqa}; some examples include $\rho\rightarrow\pi\gamma^*, B\rightarrow\rho+\ell\ell,$ and $B\rightarrow K^*+\ell\ell\rightarrow K\pi+\ell\ell$

As mentioned above, this result assumes that the particle involved have no intrinsic spin. In order to shed some light onto how systems with nonzero spin can be studied using this formalism, Agadjanov et al. investigated what would be needed to determine the $N\gamma^*\rightarrow\Delta\rightarrow N\pi$ form factors via lattice QCD~\cite{Agadjanov:2014kha}. In their work, they assumed the final two-particle state to be at rest and ignored effects due to partial wave mixing. Nevertheless, they demonstrate that infinite volume transition amplitudes, or equivalently form factors, for this process can in principal be determined. In the limits considered in the aforementioned reference, the result agree with Eq.~\ref{eq:matJaieps}.
%%%%%%%%%%%%%%%%%%%%%%%%%%%%%%%%%%%%%%%%%%%%%%%%%%%%%%%%%%%%%%%%%%%%%
%%%%%%%%%%%%%%%%%%%%%%%%%%%%%%%%%%%%%%%%%%%%%%%%%%%%%%%%%%%%%%%%%%%%%
\subsection{Elastic/inelastic form factors of unstable states}
Being able to generalize this framework for $\textbf{2}\rightarrow\textbf{2}$ would be a significant first step towards being able to study, for example, parity violation in proton-proton scattering or form factors of hadronic resonances. Yet, the generalization of this formalism is challenging. Currently, no parametrization-independent relation between the finite volume matrix elements and infinite volume amplitudes has been found, e.g., see Refs.~\cite{Detmold:2004qn, Bernard:2012bi, Briceno:2012yi}.

%%%%%%%%%%%%%%%%%%%%%%%%%%%%%%%%%%%%%%%%%%%%%%%%%%%%%%%%%%%%%%%%%%%%%
\subsection{Long range contribution to electroweak processes via lattice QCD}
Lattice QCD calculations of weak decays used to determine the CKM matrix element, have reached a level of precision (see Ref.~\cite{Aoki:2013ldr} for a recent review on the topic) where long range contributions to these must be incorporated. Two prime examples where these effects are expected to play an important role are the $K_L-K_S$ mass splitting, $\Delta m_K$, and the CP violating $K^0-\bar{K}^0$ mixing parameter, $\epsilon_K$. For the $K_L-K_S$ mass splitting, these contributions can be understood as a weak current annihilating the incoming kaon state and creating a two or three pion state that propagates for arbitrary long distances and is then turned onto a final kaon via the weak interaction. Performing this calculation in a finite volume would result in power-law finite volume artifacts associated with the intermediate two/three pion states going on-shell and sampling the boundaries of the volume. Therefore, the challenges associated with the studies of transition amplitudes discussed above, are closely related to those regarding the studies of nonlocal contribution to electroweak matrix elements. The first formal attempts towards estimating these large volume have been recently performed~\cite{Christ:2010gi, Christ:2014qaa}. In these works, it is assumed that the three pion contribution can be ignored and that the only contributing partial wave is the S-wave. Within these approximations, the authors find that he finite volume $K_L-K_S$ mass splitting, $\Delta m_K^{L}$, can be written in terms of the infinite volume splitting, $\Delta m_K$, up to an additive finite volume function,  
\begin{align}
\Delta m_K^{L}&=\Delta m_K-2\pi~\langle\bar{K}^0;L|H_{W,L}|{\pi\pi,n_0};L\rangle
 \langle {\pi\pi,n_0};L|H_{W,L}|\bar{K}^0;L\rangle
\left.\left[\cot(\pi h)\frac{\partial h}{\partial E}\right]\right|_{E=m_K},
\label{eq:mKL}
\end{align}
where $h(E,L)\equiv\phi(E,L)+\delta_{0,0}(E)$, $\phi$ is a kinematic function related to the Zeta functions~\cite{Luscher:1990ux} and $\delta_{0,0}$ is the isosinglet, S-wave $\pi\pi$ phase shift. $H_{W,L}$ is the finite volume weak Hamiltonian, and $|{\pi\pi,n_0};L\rangle$ denotes the ground state ``$\pi\pi$'' ground state evaluated at the kaon mass. Therefore, in order to reliably determine $\Delta m_K$ one must first determine the $\pi\pi$ spectrum and use Eq.~\ref{eq:QC} to constrain $\delta_{0,0}$ and its derivative. Furthermore, one would need to determine $\langle\bar{K}^0;L|H_{W,L}|{\pi\pi,n_0};L\rangle$ and $\langle {\pi\pi,n_0};L|H_{W,L}|\bar{K}^0;L\rangle$.

 Currently there are two lattice calculations of the long distance contribution to $K_L-K_S$ mass splitting have been performed~\footnote{The first determination was performed using $m_\pi\sim421~$MeV~\cite{Christ:2012se}.}. The most recent used $m_\pi=330~$MeV, $m_K=575~$MeV, a lattice spacing of $1/a = 1.729(28)$~GeV and a volume of $24^3 \times 64$ in lattice units~\cite{Bai:2014cva}. For these lattices, the kaon resides below the two pion threshold. From Eq.~\ref{eq:mKL} one can show that the finite volume effects are expected to be exponentially suppressed with the volume and can therefore be safely neglected. % $\sim\mathcal{O}(e^{-\kappa L})$ where $\kappa\sim160$~MeV is the binding momentum of the two-pion state at the kaon mass and for the volumes used this would result in $\kappa L\sim2.3$. 
 For these values of the masses and lattice spacing, Bai et al. obtain $\Delta m_K = 3.19(41)(96)\times10^{-12}~MeV$, which is surprisingly good agreement with the experimental value of $\Delta m_K=3.483(6)\times10^{-12}~MeV$~\cite{Agashe:2014kda}.~\footnote{For a recent exploration of the systematics associated with this calculation at $m_\pi~\sim171$~MeV see Ref.~\cite{Bai:2014hta}.} As calculations are performed closer to the physical point, the finite volume effects addressed by Eq.~\ref{eq:mKL} will need to be taken into account and even the effects due to intermediate three pion states will need to be considered. This would require first obtaining the generalization of Eq.~\ref{eq:mKL} that incorporate intermediate three-particle states as dictated by the three-body formalism review in Sec.~\ref{sec:3body} or its extensions to come.

\section{Outlook of the field}

Lattice QCD has proven to be a remarkably powerful tool to study low to medium energy hadronic physics. Yet the power of this tool strongly depends on our capability to understand its numerical output. As discussed extensively here, the interpretation of observables involving few-body systems is a subtle one. There has been a great deal of progress towards developing the ``\emph{theoretical infrastructure}'' to be able to perform lattice QCD calculations of said systems, and there is a ``\emph{universal}'' picture emerging for relating finite volume quantities to the physical observables. 

At this point in time, it is fair to say that following are observables that are formally and rigorously understood: \emph{two-body coupled-channel spectrum with arbitrary spin} and \emph{$\textbf{1}\rightarrow\textbf{2}$ matrix elements without intrinsic spin}. Meanwhile the following observables are just now beginning to be understood:
\\
\indent \emph{1. Finite volume spectrum involving three or more particles, }
\\
\indent \emph{2. matrix elements for $\textbf{1}\rightarrow\textbf{2}$, $\textbf{2}\rightarrow\textbf{2}$, $\textbf{1}\rightarrow\textbf{3}$, $\ldots$ processes with arbitrary spin, }
\\
\indent \emph{3. long range contributions to matrix elements involving two or more intermediate particles, }
\\
\indent \emph{4. non-perturbative QED effects.}
\\\noindent
It is safe to expect a great deal more of formal developments on these and many more few-body quantities. In closing, I would like to re-emphasize that we are currently developing a user manual of sorts that will allow us to extend the capabilities of Lattice QCD toward more interesting and complex systems. As we explore the frontier of the field we should not expect precision but rather we should expect the unexpected.~\footnote{For extended overviews of the current status and outlook of the field, I point the reader to Refs.~\cite{Walker-Loud:2014iea, Briceno:2014tqa, Beane:2014oea, Prelovsek:2014zga}.}

\section*{Acknowledgements}
\noindent
RB acknowledges support from the U.S. Department of Energy contract DE-AC05-06OR23177, under which Jefferson Science Associates, LLC, manages and operates the Jefferson Lab. RB would like to thank Kostas Orginos, Robert Edwards, Andr\'e Walker-Loud, Maxwell Hansen, Christopher Monahan, David Wilson and Zohreh Davoudi for their feedback in preparation to this talk. Also, RB would like to thank Jozef Dudek, Andr\'e Walker-Loud, and David Wilson for permitting to reproduce figures from their previous work. 

\bibliographystyle{iopart-num}
\bibliography{bibi}

\providecommand{\newblock}{}
\begin{thebibliography}{10}
\expandafter\ifx\csname url\endcsname\relax
  \def\url#1{{\tt #1}}\fi
\expandafter\ifx\csname urlprefix\endcsname\relax\def\urlprefix{URL }\fi
\providecommand{\eprint}[2][]{\url{#2}}
% Bibliography created with iopart-num v2.1
% /biblio/bibtex/contrib/iopart-num

\bibitem{Dudek:2012xn}
Dudek J~J, Edwards R~G and Thomas C~E 2013 {\em Phys.Rev.\/} {\bf D87} 034505
  (\textit{Preprint} \eprint{1212.0830})

\bibitem{Beane:2012vq}
Beane S~R, Chang E, Cohen S~D, Detmold W, Lin H {\em et~al.\/} 2013 {\em
  Phys.Rev.\/} {\bf D87} 034506 (\textit{Preprint} \eprint{1206.5219})

\bibitem{Aaij:2013qta}
Aaij R {\em et~al.\/} (LHCb collaboration) 2013 {\em Phys.Rev.Lett.\/} {\bf
  111} 191801 (\textit{Preprint} \eprint{1308.1707})

\bibitem{Detmold:2008fn}
Detmold W, Savage M~J, Torok A, Beane S~R, Luu T~C {\em et~al.\/} 2008 {\em
  Phys.Rev.\/} {\bf D78} 014507 (\textit{Preprint} \eprint{0803.2728})

\bibitem{Thomas:2011rh}
Thomas C~E, Edwards R~G and Dudek J~J 2012 {\em Phys.Rev.\/} {\bf D85} 014507
  (\textit{Preprint} \eprint{1107.1930})

\bibitem{Basak:2005ir}
Basak S {\em et~al.\/} (Lattice Hadron Physics Collaboration (LHPC)) 2005 {\em
  Phys.Rev.\/} {\bf D72} 074501 (\textit{Preprint} \eprint{hep-lat/0508018})

\bibitem{Peardon:2009gh}
Peardon M {\em et~al.\/} (Hadron Spectrum Collaboration) 2009 {\em Phys.Rev.\/}
  {\bf D80} 054506 (\textit{Preprint} \eprint{0905.2160})

\bibitem{Dudek:2010wm}
Dudek J~J, Edwards R~G, Peardon M~J, Richards D~G and Thomas C~E 2010 {\em
  Phys.Rev.\/} {\bf D82} 034508 (\textit{Preprint} \eprint{1004.4930})

\bibitem{Edwards:2011jj}
Edwards R~G, Dudek J~J, Richards D~G and Wallace S~J 2011 {\em Phys.Rev.\/}
  {\bf D84} 074508 (\textit{Preprint} \eprint{1104.5152})

\bibitem{Dudek:2012ag}
Dudek J~J and Edwards R~G 2012 {\em Phys.Rev.\/} {\bf D85} 054016
  (\textit{Preprint} \eprint{1201.2349})

\bibitem{Dudek:2012gj}
Dudek J~J, Edwards R~G and Thomas C~E 2012 {\em Phys.Rev.\/} {\bf D86} 034031
  (\textit{Preprint} \eprint{1203.6041})

\bibitem{Morningstar:2013bda}
Morningstar C, Bulava J, Fahy B, Foley J, Jhang Y {\em et~al.\/} 2013 {\em
  Phys.Rev.\/} {\bf D88} 014511 (\textit{Preprint} \eprint{1303.6816})

\bibitem{Lang:2013uwa}
Lang C and Verduci V 2014 {\em Int.J.Mod.Phys.Conf.Ser.\/} {\bf 26} 1460056
  (\textit{Preprint} \eprint{1309.4677})

\bibitem{Detmold:2010au}
Detmold W and Savage M~J 2010 {\em Phys.Rev.\/} {\bf D82} 014511
  (\textit{Preprint} \eprint{1001.2768})

\bibitem{Detmold:2013gua}
Detmold W and Nicholson A~N 2013 {\em Phys.Rev.\/} {\bf D88} 074501
  (\textit{Preprint} \eprint{1308.5186})

\bibitem{Doi:2012xd}
Doi T and Endres M~G 2013 {\em Comput. Phys. Commun.\/} {\bf 184} 117
  (\textit{Preprint} \eprint{1205.0585})

\bibitem{Detmold:2012eu}
Detmold W and Orginos K 2013 {\em Phys.Rev.\/} {\bf D87} 114512
  (\textit{Preprint} \eprint{1207.1452})

\bibitem{Lepage89}
Lepage G~P 1989  Invited lectures given at TASI'89 Summer School, Boulder, CO,
  Jun 4-30, 1989

\bibitem{Grabowska:2012ik}
Grabowska D, Kaplan D~B and Nicholson A~N 2013 {\em Phys.Rev.\/} {\bf D87}
  014504 (\textit{Preprint} \eprint{1208.5760})

\bibitem{Endres:2011jm}
Endres M~G, Kaplan D~B, Lee J~W and Nicholson A~N 2011 {\em Phys.Rev.Lett.\/}
  {\bf 107} 201601 (\textit{Preprint} \eprint{1106.0073})

\bibitem{Detmold:2014hla}
Detmold W and Endres M~G 2014 {\em Phys.Rev.\/} {\bf D90} 034503
  (\textit{Preprint} \eprint{1404.6816})

\bibitem{Detmold:2014rfa}
Detmold W and Endres M~G 2014  (\textit{Preprint} \eprint{1409.5667})

\bibitem{Maiani:1990ca}
Maiani L and Testa M 1990 {\em Phys.Lett.\/} {\bf B245} 585--590

\bibitem{Briceno:2014uqa}
Briceno R~A, Hansen M~T and Walker-Loud A 2014  (\textit{Preprint}
  \eprint{1406.5965})

\bibitem{BHWL}
Briceno R~A, Hansen M~T and Walker-Loud A 2014

\bibitem{Luscher:1986pf}
Luscher M 1986 {\em Commun.Math.Phys.\/} {\bf 105} 153--188

\bibitem{Luscher:1990ux}
Luscher M 1991 {\em Nucl.Phys.\/} {\bf B354} 531--578

\bibitem{Lellouch:2000pv}
Lellouch L and Luscher M 2001 {\em Commun.Math.Phys.\/} {\bf 219} 31--44
  (\textit{Preprint} \eprint{hep-lat/0003023})

\bibitem{Kim:2005gf}
Kim C, Sachrajda C and Sharpe S~R 2005 {\em Nucl.Phys.\/} {\bf B727} 218--243
  (\textit{Preprint} \eprint{hep-lat/0507006})

\bibitem{Christ:2005gi}
Christ N~H, Kim C and Yamazaki T 2005 {\em Phys.Rev.\/} {\bf D72} 114506
  (\textit{Preprint} \eprint{hep-lat/0507009})

\bibitem{Rummukainen:1995vs}
Rummukainen K and Gottlieb S~A 1995 {\em Nucl. Phys.\/} {\bf B450} 397--436
  (\textit{Preprint} \eprint{hep-lat/9503028})

\bibitem{Bernard:2008ax}
Bernard V, Lage M, Meissner U~G and Rusetsky A 2008 {\em JHEP\/} {\bf 0808} 024
  (\textit{Preprint} \eprint{0806.4495})

\bibitem{Hansen:2012tf}
Hansen M~T and Sharpe S~R 2012 {\em Phys.Rev.\/} {\bf D86} 016007
  (\textit{Preprint} \eprint{1204.0826})

\bibitem{Briceno:2012yi}
Briceno R~A and Davoudi Z 2013 {\em Phys. Rev. D. 88,\/} {\bf 094507} 094507
  (\textit{Preprint} \eprint{1204.1110})

\bibitem{Briceno:2014oea}
Briceno R~A 2014 {\em Phys.Rev.\/} {\bf D89} 074507 (\textit{Preprint}
  \eprint{1401.3312})

\bibitem{Polejaeva:2012ut}
Polejaeva K and Rusetsky A 2012 {\em Eur.Phys.J.\/} {\bf A48} 67
  (\textit{Preprint} \eprint{1203.1241})

\bibitem{Briceno:2012rv}
Briceno R~A and Davoudi Z 2012 {\em Phys.Rev.\/} {\bf D87} 094507
  (\textit{Preprint} \eprint{1212.3398})

\bibitem{Hansen:2014lya}
Hansen M~T and Sharpe S~R 2014  (\textit{Preprint} \eprint{1409.7012})

\bibitem{Hansen:2014eka}
Hansen M~T and Sharpe S~R 2014  (\textit{Preprint} \eprint{1408.5933})

\bibitem{Guo:2013qla}
Guo P 2013  (\textit{Preprint} \eprint{1303.3349})

\bibitem{Kreuzer:2010ti}
Kreuzer S and Hammer H~W 2011 {\em Phys.Lett.\/} {\bf B694} 424--429
  (\textit{Preprint} \eprint{1008.4499})

\bibitem{Dudek:2014qha}
Dudek J~J, Edwards R~G, Thomas C~E and Wilson D~J 2014  (\textit{Preprint}
  \eprint{1406.4158})

\bibitem{Bernard:2012bi}
Bernard V, Hoja D, Meissner U~G and Rusetsky A 2012 {\em JHEP\/} {\bf 1209} 023
  (\textit{Preprint} \eprint{1205.4642})

\bibitem{Guo:2013vsa}
Guo P 2013 {\em Phys.Rev.\/} {\bf D88} 014507 (\textit{Preprint}
  \eprint{1304.7812})

\bibitem{Blum:2007cy}
Blum T, Doi T, Hayakawa M, Izubuchi T and Yamada N 2007 {\em Phys.Rev.\/} {\bf
  D76} 114508 (\textit{Preprint} \eprint{0708.0484})

\bibitem{Blum:2010ym}
Blum T, Zhou R, Doi T, Hayakawa M, Izubuchi T {\em et~al.\/} 2010 {\em
  Phys.Rev.\/} {\bf D82} 094508 (\textit{Preprint} \eprint{1006.1311})

\bibitem{Aoki:2012st}
Aoki S, Ishikawa K, Ishizuka N, Kanaya K, Kuramashi Y {\em et~al.\/} 2012 {\em
  Phys.Rev.\/} {\bf D86} 034507 (\textit{Preprint} \eprint{1205.2961})

\bibitem{deDivitiis:2013xla}
de~Divitiis G, Frezzotti R, Lubicz V, Martinelli G, Petronzio R {\em et~al.\/}
  2013 {\em Phys.Rev.\/} {\bf D87} 114505 (\textit{Preprint}
  \eprint{1303.4896})

\bibitem{Borsanyi:2014jba}
Borsanyi S, Durr S, Fodor Z, Hoelbling C, Katz S {\em et~al.\/} 2014
  (\textit{Preprint} \eprint{1406.4088})

\bibitem{Duncan:1996xy}
Duncan A, Eichten E and Thacker H 1996 {\em Phys.Rev.Lett.\/} {\bf 76}
  3894--3897 (\textit{Preprint} \eprint{hep-lat/9602005})

\bibitem{Hayakawa:2008an}
Hayakawa M and Uno S 2008 {\em Prog.Theor.Phys.\/} {\bf 120} 413--441
  (\textit{Preprint} \eprint{0804.2044})

\bibitem{Davoudi:2014qua}
Davoudi Z and Savage M~J 2014 {\em Phys.Rev.\/} {\bf D90} 054503
  (\textit{Preprint} \eprint{1402.6741})

\bibitem{Beane:2014qha}
Beane S~R and Savage M~J 2014  (\textit{Preprint} \eprint{1407.4846})

\bibitem{Lee:2014iha}
Lee J~W and Tiburzi B~C 2014  (\textit{Preprint} \eprint{1407.8159})

\bibitem{Beane:2014ora}
Beane S, Chang E, Cohen S, Detmold W, Lin H {\em et~al.\/} 2014
  (\textit{Preprint} \eprint{1409.3556})

\bibitem{Detmold:2004qn}
Detmold W and Savage M~J 2004 {\em Nucl.Phys.\/} {\bf A743} 170--193
  (\textit{Preprint} \eprint{hep-lat/0403005})

\bibitem{Agadjanov:2014kha}
Agadjanov A, Bernard V, Meißner U~G and Rusetsky A 2014 {\em Nucl.Phys.\/} {\bf
  B886} 1199--1222 (\textit{Preprint} \eprint{1405.3476})

\bibitem{Ishizuka:2014nfa}
Ishizuka N, Ishikawa K, Ukawa A and Yoshié T 2014  (\textit{Preprint}
  \eprint{1410.8237})

\bibitem{Blum:2012uk}
Blum T, Boyle P, Christ N, Garron N, Goode E {\em et~al.\/} 2012 {\em
  Phys.Rev.\/} {\bf D86} 074513 (\textit{Preprint} \eprint{1206.5142})

\bibitem{Boyle:2012ys}
Boyle P {\em et~al.\/} (RBC, UKQCD) 2013 {\em Phys.Rev.Lett.\/} {\bf 110}
  152001 (\textit{Preprint} \eprint{1212.1474})

\bibitem{Blum:2011pu}
Blum T, Boyle P, Christ N, Garron N, Goode E {\em et~al.\/} 2011 {\em
  Phys.Rev.\/} {\bf D84} 114503 (\textit{Preprint} \eprint{1106.2714})

\bibitem{Blum:2011ng}
Blum T, Boyle P, Christ N, Garron N, Goode E {\em et~al.\/} 2012 {\em
  Phys.Rev.Lett.\/} {\bf 108} 141601 (\textit{Preprint} \eprint{1111.1699})

\bibitem{Aoki:2013ldr}
Aoki S, Aoki Y, Bernard C, Blum T, Colangelo G {\em et~al.\/} 2014 {\em
  Eur.Phys.J.\/} {\bf C74} 2890 (\textit{Preprint} \eprint{1310.8555})

\bibitem{Christ:2010gi}
Christ N~H (RBC Collaboration, UKQCD Collaboration) 2010  (\textit{Preprint}
  \eprint{1012.6034})

\bibitem{Christ:2014qaa}
Christ N, Martinelli G and Sachrajda C 2014 {\em PoS\/} {\bf LATTICE2013} 399
  (\textit{Preprint} \eprint{1401.1362})

\bibitem{Christ:2012se}
Christ N, Izubuchi T, Sachrajda C, Soni A and Yu J (RBC and UKQCD
  Collaborations) 2013 {\em Phys.Rev.\/} {\bf D88} 014508 (\textit{Preprint}
  \eprint{1212.5931})

\bibitem{Bai:2014cva}
Bai Z, Christ N, Izubuchi T, Sachrajda C, Soni A {\em et~al.\/} 2014 {\em
  Phys.Rev.Lett.\/} {\bf 113} 112003 (\textit{Preprint} \eprint{1406.0916})

\bibitem{Agashe:2014kda}
Olive K {\em et~al.\/} (Particle Data Group) 2014 {\em Chin.Phys.\/} {\bf C38}
  090001

\bibitem{Bai:2014hta}
Bai Z 2014  (\textit{Preprint} \eprint{1411.3210})

\bibitem{Walker-Loud:2014iea}
Walker-Loud A 2014 {\em PoS\/} {\bf LATTICE2013} 013 (\textit{Preprint}
  \eprint{1401.8259})

\bibitem{Briceno:2014tqa}
Briceno R~A, Davoudi Z and Luu T~C 2014  (\textit{Preprint} \eprint{1406.5673})

\bibitem{Beane:2014oea}
Beane S~R, Detmold W, Orginos K and Savage M~J 2014  (\textit{Preprint}
  \eprint{1410.2937})

\bibitem{Prelovsek:2014zga}
Prelovsek S 2014  (\textit{Preprint} \eprint{1411.0405})

\end{thebibliography}

\end{document}